\documentclass[12pt]{article}
\begin{document}
\title{{\bf Symmetric-Bounce Quantum State of the Universe}
\thanks{Alberta-Thy-09-09, arXiv:0907.1893}}
\author{
Don N. Page
\thanks{Internet address:
don@phys.ualberta.ca}
\\
Theoretical Physics Institute\\
Department of Physics, University of Alberta\\
Room 238 CEB, 11322 -- 89 Avenue\\
Edmonton, Alberta, Canada T6G 2G7
}
\date{(2009 August 14)}

\maketitle
\large
\begin{abstract}
\baselineskip 18 pt

	A proposal is made for the quantum state of the universe that
has an initial state that is macroscopically time symmetric about a
homogeneous, isotropic bounce of extremal volume and that at that
bounce is microscopically in the ground state for inhomogeneous and/or
anisotropic perturbation modes.  The coarse-grained entropy is minimum
at the bounce and then grows during inflation as the modes become
excited away from the bounce and interact (assuming the presence of an
inflaton, and in the part of the quantum state in which the inflaton
is initially large enough to drive inflation).  The part of this pure
quantum state that dominates for observations is well approximated by
quantum processes occurring within a Lorentzian expanding macroscopic
universe.  Because this part of the quantum state has no negative
Euclidean action, one can avoid the early-time Boltzmann brains and
Boltzmann solar systems that appear to dominate observations in the
Hartle-Hawking no-boundary wavefunction.

\end{abstract}
\normalsize

\baselineskip 13.9 pt

\newpage 

\section*{Introduction}

	Even if physicists succeed in finding a so-called `Theory of
Everything' or TOE that gives the full set of dynamical laws for our
universe, it appears that that will be insufficient to explain our
past observations and to predict new ones.  The reason is that each
set of dynamical laws, at least of the kind we are familiar with,
permits a wide variety of solutions, most of which would be
inconsistent with our observations.  We need a set of initial
conditions and/or other boundary conditions to restrict the possible
solutions to fit what we observe.  In a quantum description of the
universe with fixed dynamical laws (the analogue of the
Schr\"{o}dinger equation for nonrelativistic quantum mechanics), we
need not only these dynamical laws but also the quantum state itself (cf.\ \cite{DoOur}). (We also need the rules for extracting observational probabilities from the quantum state \cite{cmwvw,insuff,brd,ba} for solving the measure problem in cosmology, which is another extremely important issue, but I shall not focus on that in this paper.)

	To put it another way, our observations strongly suggest that
our observed portion (or subuniverse \cite{Wein87} or bubble universe
\cite{Linde96,Vil97} or pocket universe \cite{Guth00}) of the entire
universe (or multiverse
\cite{James,Lodge,Leslie,Gell-Mann,Deutsch,Dyson,Rees,Carr} or
metauniverse \cite{Vil95} or omnium \cite{Penrose-book} or megaverse
\cite{Susskind-book}) is much more special than is implied purely by
the known dynamical laws.  For example, it is seen to be enormously
larger than the Planck scale, with small large-scale curvature, and
with approximate homogeneity and isotropy of the matter distribution
on the largest scales that we can see today.  It especially seems to
have had extraordinarily high order in the early universe to enable
its coarse-grained entropy to increase and to give us the observed
second law of thermodynamics \cite{Tolman,Davies,Penrose-in-HI}.  The known dynamical laws do not imply these observed conditions.

	Leading proposals for special quantum states of the universe
have been the Hartle-Hawking `no-boundary' proposal
\cite{HVat,HH,H,HalHaw,P,Hal,HLL,HH1,Page-Hawking-BD,HH2,HHH} and the
`tunneling' proposals of Vilenkin, Linde, and others
\cite{V,TL,TZS,TR,TVV,TGV}. In simplified toy models with a suitable
inflaton, both of these classes of models have seemed to lead to the
special observed features of our universe noted above.

	However, Leonard Susskind \cite{Susspriv} (cf.\
\cite{DKS,GKS,Sus03}) has made the argument, which I have elaborated
\cite{DP2006}, that in the no-boundary proposal the cosmological
constant or quintessence or dark energy that is the source of the
present observations of the cosmic acceleration
\cite{Perl,Riess,PTW,Tonry,WMAP,Tegmark,Astier} would give a very large
Euclidean 4-hemisphere as an extremum of the Hartle-Hawking path
integral that would apparently swamp the extremum from rapid early
inflation by amplitude factors of the order of $e^{10^{122}}$. 
Therefore, to very high probability, the present universe should be very
nearly empty de Sitter spacetime, which is certainly not what we
observe.  Even if we restrict to the very rare cases in which a solar
system like ours occurs, the probability in the Hartle-Hawking
no-boundary proposal seems to be much, much higher for a single solar
system in an otherwise empty universe than for a solar system surrounded
by other stars such as what we observe.

	The tunneling proposals have also been criticized for various
problems \cite{BH,TGV,Lin98,HT,TH,V98}.  For example, the main
difference from the Hartle-Hawking no-boundary proposal seems to be the
sign of the Euclidean action \cite{V,TL}.  It then seems problematic to
take the opposite sign for inhomogeneous and/or anisotropic
perturbations without leading to some instabilities, and it is not clear
how to give a sharp distinction between the modes that are supposed to
have the reversed sign of the action and the modes that are supposed to
retain the usual sign of the action.  Vilenkin and his collaborators have emphasized \cite{V,TVV,TGV} that the instabilities do not seem to apply to his particular tunneling proposal, which does not just reverse the sign of the Euclidean action.  However, Vilenkin (with Garriga) admits \cite{TGV} that ``both wavefunctions are far from being rigorous
mathematical objects with clearly specified calculational procedures. 
Except in the simplest models, the actual calculations of
$\psi_{\mathrm{T}}$ and $\psi_{\mathrm{HH}}$ involve additional
assumptions which appear reasonable, but are not really well
justified.''

	Therefore, at least unless and until any of these proposals can
be made rigorous and can be shown conclusively to avoid the problems
attributed to them, it is worth searching for and examining other
possibilities for the quantum state of the universe or multiverse.  In a
previous paper \cite{nb}, I proposed a `no-bang' quantum state which is
the equal mixture of the Giddings-Marolf states \cite{GM} that are
asymptotically single de Sitter spacetimes in both past and future and
are regular on the throat or neck of minimal three-volume.  However, it
does not appear to work if one adopts my proposal of volume averaging
\cite{cmwvw} to help solve the late-time aspect of the Boltzmann brain
problem.

	The Boltzmann brain problem \cite{DKS,Albrecht,AS,Page05,YY,
Page06a,BF,Page06b,Linde06,Page06c,Vil06,Page06d,Vanchurin,Banks,Carlip,
HS,GM,Giddings,typdef,LWc,DP07,Bou08,BFYa,ADSV,Gott,typder,FL} is the
problem that many cosmological theories seem to predict that our
observations would be highly improbable in comparison with much more
disordered observations of Boltzmann brains that these theories predict
should enormously dominate over ordinary observers.  Boltzmann brains
are observers that appear from thermal or vacuum fluctuations.  The
probability of a Boltzmann brain per four-volume is extremely tiny (say
roughly $e^{-10^{42}}$ \cite{Page06b,Page06c,Page06d}), but if the
universe lasts for an infinite time, and especially if its three-volume
grows asymptotically exponentially, and if there are only a finite
number of ordinary observers per comoving three-volume, then per
comoving volume the Boltzmann brains will dominate and make our ordered
observations very atypical and improbable relative to the much more
disordered typical Boltzmann brain observations.  (The dominance by
Boltzmann brains at very late times, which might occur in any universe
that lasts forever, I call the late-time Boltzmann brain problem; the
Hartle-Hawking no-boundary proposal appears to suffer from what might be
called an early-time Boltzmann brain problem, that at all times Boltzmann
brains seem to dominate over ordinary observers \cite{DKS,GKS,Sus03,
Susspriv,DP2006}.)

	Originally I proposed a solution to the Boltzmann brain problem
in which the universe might be likely to decay before Boltzmann brains
would dominate \cite{Page05,Page06a,Page06b,Page06c,Page06d}, but this
seemed to require fine-tuning of whatever physics might determine the
decay rate (though see \cite{Page09} for a possible anthropic
explanation of this decay rate).  Therefore, I turned to another
possible solution, that one should go from volume weighting to volume
averaging \cite{cmwvw} to extract observational probabilities.  This
would eliminate the effect of the exponentially growing 3-volumes in the
asymptotic future, though there still remains a much less rapid
divergence on the weighting of Boltzmann brains from an infinite future
lifetime of the universe, unless one went beyond 3-volume averaging to
4-volume averaging that would allow a possible anthropic explanation of
a decaying universe \cite{Page09}.  However, if one goes from volume
weighting to volume averaging to mitigate the late-time Boltzmann brain
problem, the no-bang state then appears to suffer qualitatively from the
same problem as the no-boundary state of being dominated by thermal
perturbations of nearly empty de Sitter spacetime, so that almost all
observers would presumably be Boltzmann brains.  Since this would almost
certainly make our observations very unlikely, the no-bang proposal
apparently is observationally excluded if one uses volume averaging
rather than volume weighting.  (The no-boundary state appears to be
excluded if either rule were used for extracting probabilities from the
quantum state, since it has both an early-time and a late-time Boltzmann
brain problem.)

	In this paper, instead of the mixed `no-bang' state, I shall
propose a pure quantum state in which the Giddings-Marolf seed state
\cite{GM} (before group averaging over diffeomorphisms) consists of
quantum fluctuations about a uniform superposition of Lorentzian macroscopic components that are each time symmetric about a bounce of extremal 3-volume, with the quantum fluctuations being in their ground state at that moment of time symmetry for the macroscopic 4-geometry.  With both signs of the Lorentzian time away from this momentarily-static bounce, the 3-volume will expand, typically in an inflationary manner if the matter is dominated by a sufficiently large homogeneous component of a scalar inflaton field.  This inflationary expansion will then produce
parametric amplification of the inhomogeneous and anisotropic modes in
the usual manner to give density fluctuations at the end of inflation
that then grow gravitationally to become nonlinear and produce the
structure that we observe.

	A slight aesthetic disadvantage of the symmetric-bounce quantum
state in comparison with the no-boundary state is that in the
symmetric-bounce proposal, the inhomogeneous fluctuations are put into
their ground state at the bounce by a part of the proposal that is
logically separate from the part of the proposal that gives the
behavior of the homogeneous modes, whereas in the no-boundary proposal
the behavior of both the inhomogeneous and homogeneous modes come out
together from the same part of that proposal, that the histories that
contribute to the path integral are regular on a complete complexified
Euclidean manifold with no boundary other than the one on which the
wavefunction is evaluated.  However, this seems to be a small price to
pay for avoiding the huge negative Euclidean actions of many
nearly-empty de Sitter histories in the no-boundary proposal that make
nearly empty spacetime much more probable than a nearly
Friedmann-Robertson-Walker spacetime with high densities at early
times that would fit our observations much better.  To avoid making our observation of distant stars extremely improbable, as it appears to be in the no-boundary proposal, it seems well worth giving up the simple no-boundary unified description of the behavior of both the homogeneous inflationary modes and the inhomogeneous fluctuation modes.

\section{Homogeneous modes with an inflaton and a\\
cosmological constant}

	First, let us focus on the behavior of the homogeneous,
isotropic modes of the symmetric-bounce quantum seed state.  That is,
take each quasiclassical component of the macroscopic spacetime geometry, without the quantum fluctuations, to be a Friedmann-Robertson-Walker (FRW) model driven by homogeneous matter fields.  For concreteness and simplicity, consider the case of a positive cosmological constant $\Lambda = 3/b^2$ and a single inflaton that is a homogeneous free scalar field $\phi(t)$ of mass $m$, and take the FRW model to be $k=+1$ so that the spatial sections are homogeneous, isotropic 3-spheres of radius $a(t)$.  Then the macroscopic spacetime metric can be taken to be
\begin{equation}
ds^2 = -N^2dt^2 + a^2(t)d\Omega_3^2.
\label{eq:a}
\end{equation}

	Using units in which $\hbar = c = 1$, but writing $G$
explicitly, one can write the Lorentzian action as
(cf.\ \cite{dnpcqg})
\begin{eqnarray}
S &=& \int Ndt 2\pi^2 a^3 \left\{\frac{3}{8\pi G}
        \left[-\left(\frac{1}{Na}\frac{da}{dt}\right)^2
	      +\frac{1}{a^2}-\frac{\Lambda}{3}\right]
	+\frac{1}{2}\left(\frac{1}{N}\frac{d\phi}{dt}\right)^2
	-\frac{1}{2}m^2\phi^2\right\} \nonumber \\
  &=& \frac{3\pi}{4G}\int Ndt a^3 \left\{
        -\left(\frac{1}{Na}\frac{da}{dt}\right)^2
	+\left(\frac{1}{N}\frac{d\varphi}{dt}\right)^2
	+\frac{1}{a^2}-\frac{1}{b^2}-m^2\varphi^2\right\} \nonumber \\
  &=& \frac{3\pi}{4Gm^2}\int ndt r^3 \left\{
        -\left(\frac{1}{nr}\frac{dr}{dt}\right)^2
	+\left(\frac{1}{n}\frac{d\varphi}{dt}\right)^2
	+\frac{1}{r^2}-\lambda-\varphi^2\right\} \nonumber \\
  &=& \frac{3\pi}{4}\frac{m_{\mathrm{Pl}}^2}{m^2}\int ndt e^{3\alpha}
       \left[-n^{-2}(\dot{\alpha}^2-\dot{\varphi}^2)
        +e^{-2\alpha}-\lambda-\varphi^2\right] \nonumber \\
  &=& \frac{1}{2}\int ndt 
       \left[\left(\frac{1}{n}\frac{ds}{dt}\right)^2-V\right]
      = \frac{1}{2}\int dt 
         \left[\frac{1}{\nu}\left(\frac{d\hat{s}}{dt}\right)^2
                          -\nu\right],
\label{eq:b}
\end{eqnarray}
where $b\equiv \sqrt{3/\Lambda}$ is the radius of the throat of pure de
Sitter with the same value of the cosmological constant, $n \equiv mN$
is a rescaled lapse function that is dimensionless if $t$ is taken to be
dimensionless, $\lambda \equiv \Lambda/(3m^2) \equiv 1/(mb)^2$ is a
dimensionless measure of the cosmological constant in units given by the
mass of the inflaton, $r \equiv e^{\alpha} \equiv ma$ and $\varphi
\equiv \sqrt{4\pi G/3}\phi$ are dimensionless forms of the scale factor
and inflaton scalar field (leaving $G \equiv m_{\mathrm{Pl}}^{-2}$ to
have the dimensions of inverse mass squared or of area), an overdot
represents a derivative with respect to $t$, the DeWitt metric \cite{DeW} on the minisuperspace is
\begin{equation}
ds^2 = \frac{3\pi}{2Gm^2}e^{3\alpha}(-d\alpha^2 + d\varphi^2),
\label{eq:c}
\end{equation}
the `potential' on the minisuperspace is
\begin{equation}
V = \frac{3\pi}{2Gm^2}e^{3\alpha}(\varphi^2+\lambda-e^{-2\alpha}),
\label{eq:d}
\end{equation}
the rescaled lapse function is $\nu \equiv nV = mNV$, and the
conformal minisuperspace metric is
\begin{equation}
d\hat{s}^2 = Vds^2 = \left(\frac{3\pi}{2Gm^2}\right)^2 e^{6\alpha}
       (\varphi^2+\lambda-e^{-2\alpha})(-d\alpha^2 + d\varphi^2).
\label{eq:e}
\end{equation}

	To get some reasonable numbers for the dimensionless constants
in these equations, take $\Omega_\Lambda = 0.72\pm 0.04$ from the
third-year WMAP results of \cite{WMAP} and $H_0 = 72\pm 8$ km/s/Mpc
from the Hubble Space Telescope key project \cite{Freedman}, and drop
the error uncertainties to get $G\Lambda = 3\Omega_\lambda GH_0^2
\approx 3.4\times 10^{-122}$, which would give $b = \sqrt{3/\Lambda}
\approx 9.4\times 10^{60}\sqrt{G}$.  Then use the estimate that $m
\approx 1.5\times 10^{-6} G^{-1/2} \approx 7.5\times 10^{-6} (8\pi
G)^{-1/2}$ \cite{Lindebook,LL} from the measured fluctuations of the
cosmic microwave background to get that the prefactor of the action is
$(3\pi/4)(m_{\mathrm{Pl}}/m)^2 \approx 1.0\times 10^{12}$, and the
dimensionless measure of the cosmological constant is $\lambda \equiv
\Lambda/(3m^2) \equiv 1/(mb)^2 \approx 5.0\times 10^{-111}$.  Thus
$\lambda$ may be taken to be extremely tiny, and for histories in
which $\alpha$ and/or $\varphi$ are of the order of unity or greater,
the action will be very large and so should give essentially classical
behavior, at least for the homogeneous, isotropic part of the
geometry.

	The constraint equation and independent equation of motion can
now be written as
\begin{eqnarray}
 &&\left(\frac{1}{Na}\frac{da}{dt}\right)^2
=\left(\frac{1}{N}\frac{d\varphi}{dt}\right)^2
 +m^2\varphi^2 + \frac{1}{b^2} - \frac{1}{a^2}, \nonumber \\
 &&\frac{1}{N}\frac{d}{dt}\left(\frac{1}{N}\frac{d\varphi}{dt}\right)
+\left(\frac{3}{Na}\frac{da}{dt}\right)
       \left(\frac{1}{N}\frac{d\varphi}{dt}\right) + m^2\varphi^2 = 0,
\label{eq:f}
\end{eqnarray}
for general lapse function from the second form of the action above,
\begin{eqnarray}
&&\dot{r}^2 = r^2(\dot{\varphi}^2 + \varphi^2 + \lambda) - 1,
   \nonumber \\
&&\ddot{\varphi} + 3\frac{\dot{r}}{r}\dot{\varphi} + \varphi = 0,
\label{eq:g}
\end{eqnarray}
from the third form of the action with $n=1$, and
\begin{eqnarray}
&&\dot{\alpha}^2 - \dot{\varphi}^2 = \varphi^2+\lambda-e^{-2\alpha},
   \nonumber \\
&&\ddot{\varphi} + 3\dot{\alpha}\dot{\varphi} + \varphi = 0,
\label{eq:h}
\end{eqnarray}
for the fourth form of the action above with $n=1$, which will
henceforth be assumed.

Although it is a redundant equation, one may readily derive from Eqs.\
(\ref{eq:h}) that
\begin{equation}
\ddot{\alpha} = e^{-2\alpha}-3\dot{\varphi}^2 
\label{eq:h2}
\end{equation}
when $n=1$.  Then when neither side of the constraint (first) equation
part of Eqs. (\ref{eq:h}) vanishes (e.g., when $V\neq 0$), and when
$\dot{\varphi}\neq 0$, one may define $f' \equiv df/d\varphi =
\dot{f}/\dot{\varphi}$ and reduce Eqs.\ (\ref{eq:h}) to the single
second-order differential equation (cf.\ \cite{dnpcqg})
\begin{equation}
\alpha'' = \frac{(\alpha'^2-1)
            (\varphi \alpha' + 3\varphi^2 + 3\lambda - 2e^{-2\alpha})}
	    {\varphi^2+\lambda-e^{-2\alpha}}.
\label{eq:i}
\end{equation}
Alternatively, when $V\neq 0$ (or equivalently $\dot{\alpha}^2 \neq
\dot{\varphi}^2$), but when $\dot{\alpha} \neq 0$ instead of
$\dot{\varphi}\neq 0$, one can write
\begin{equation}
\frac{d^2\varphi}{d\alpha^2} = 
\frac{(d\varphi/d\alpha)^2-1}{\varphi^2+\lambda-e^{-2\alpha}}
 \left[\left(3\varphi^2 + 3\lambda - 2e^{-2\alpha}\right)
 \frac{d\varphi}{d\alpha} + \varphi\right].
\label{eq:j}
\end{equation}

Yet another way to get the equations of motion is to note that the
fifth form of the action from Eq.\ (\ref{eq:b}) gives the trajectories
of a particle of mass-squared $V$ in the DeWitt minisuperspace metric \cite{DeW} $ds^2$, and the sixth form of the action gives timelike geodesics in the conformal minisuperspace metric $d\hat{s}^2 = Vds^2$.  When one goes to the gauge $\nu = 1$, then $(d\hat{s}/dt)^2 = -1$, so that along the classical timelike geodesics of $d\hat{s}^2$, the Lorentzian action is $S = -\int dt = -\int\sqrt{-d\hat{s}^2}$, minus the proper time along the timelike geodesic of $d\hat{s}^2$.  However, one must note that the conformal metric $d\hat{s}^2 = Vds^2$ is singular at $V=0$, that is at $\varphi^2 + \lambda = e^{-2\alpha} \equiv 1/(ma)^2$, whereas there is no singularity in the DeWitt metric $ds^2$ or the spacetime metric along this hypersurface (line) in the two-dimensional minisuperspace $(\alpha,\varphi)$ under consideration.  The second-order differential equations (\ref{eq:i}) and (\ref{eq:j}) also break down at $V=0$ and must be supplemented by the continuity of $\dot{\alpha}$ and of $\dot{\varphi}$ (in a gauge in which $n \neq 0$ is continuous there) across the $V=0$ hypersurface (line).

\section{Symmetric-bounce proposal for the homogeneous modes}

My symmetric-bounce proposal for the homogeneous modes, which are
represented classically by the trajectories in the $(\alpha,\varphi)$
minisuperspace, is that one takes the set of all Lorentzian symmetric
bounce trajectories, those that have $\dot{\alpha} = \dot{\varphi} =
0$ somewhere along the classical trajectory.  By the definition Eq.\
(\ref{eq:d}) of the potential $V(\alpha,\varphi)$ and by the
constraint Eq.\ (\ref{eq:h}), this point of the trajectory will have
$V = 0$ or $a = \sqrt{3}/\sqrt{4\pi G m^2 \phi^2 + \Lambda}$ or
\begin{equation}
\alpha = \alpha_{\mathrm{bounce}}(\varphi)
      \equiv -\frac{1}{2}\ln{(\varphi^2 + \lambda)}.
\label{eq:k}
\end{equation}
The classical trajectory that has $\dot{\alpha} = \dot{\varphi} = 0$
at $(\alpha,\varphi) = (\alpha_{\mathrm{bounce}}(\varphi_b),\varphi_b)$
for some value of $\varphi_b \equiv \varphi_{\mathrm{bounce}}$ will be
time symmetric about this bounce point, so if one sets $t=0$ there and
uses a time-symmetric lapse function, $n(t) = n(-t)$, then
$(\alpha(t),\varphi(t)) = (\alpha(-t),\varphi(-t))$.

A generic trajectory in the $(\alpha,\varphi)$ minisuperspace can be
labeled by the location at which it crosses some hypersurface (e.g., at
its value of $\varphi$ on a hypersurface of fixed $\alpha$) and by its
direction there (e.g., its value of $\alpha' = d\alpha/d\varphi$), since
once the direction is fixed, the constraint equation determines the
values of both $\dot{\alpha}$ and of $\dot{\varphi}$. Thus the generic
minisuperspace trajectories form a two-parameter family.  However, the
symmetric-bounce trajectories may be labeled by the single parameter
$\varphi_b$ of the value of $\varphi$ that it has on the hypersurface
$\alpha = \alpha_{\mathrm{bounce}}(\varphi)$, since at that point on a
symmetric-bounce trajectory, the values of $\dot{\alpha}$ and of
$\dot{\varphi}$ are both determined to be zero.  Therefore, in terms of
the classical measure \cite{GHS} on the two-dimensional space of
minisuperspace trajectories, the symmetric-bounce trajectories are a set
of measure zero.  This restriction on the classical phase space of
trajectories is precisely analogous to the restriction of the
no-boundary state on the set of classical trajectories \cite{HHH},
though the details of the restriction are slightly different (precisely
real classical trajectories that have symmetric bounces for the
symmetric-bounce state).

However, since I am proposing that the quantum state is a superposition of initially quasiclassical components that give a one-parameter set of classical trajectories, to make the proposal definite I do need to give the coefficients in the quantum quantum superposition or the measure for the classical trajectories, analogous to the weighting by the exponential of minus the (negative) Euclidean action for the no-boundary proposal and by essentially the exponential of the Euclidean action for the tunneling proposal.  I shall propose that the one-parameter set of classical trajectories are uniformly distributed over the symmetric-bounce hypersurface $(\alpha_{\mathrm{bounce}}(\varphi_b),\varphi_b)$, with no weighting by the exponential of either minus or plus the Euclidean action.  Thus my symmetric-bounce quantum state has a measure that is basically the geometric mean of the no-boundary and tunneling proposals.  For such a uniform measure, $\mu(\varphi_b)d\varphi_b$, I shall take the magnitude of the metric induced on this hypersurface by the DeWitt minisuperspace metric \cite{DeW} given by Eq.\ (\ref{eq:c}), after dropping the constant factor $3\pi/(2Gm^2)$.  That is, I shall take
\begin{eqnarray}
\mu(\varphi_b)d\varphi_b 
&=& \sqrt{2Gm^2/(3\pi)}\,|ds| \nonumber \\
&=& e^{3\alpha_{\mathrm{bounce}}(\varphi_b)/2}
 \sqrt{|1-[d\alpha_{\mathrm{bounce}}(\varphi_b)/d\varphi_b]^2|}
 \,d\varphi_b
       \nonumber \\
&=& (\varphi_b^2+\lambda)^{-7/4}
     \sqrt{|\varphi_b^2+\varphi_b+\lambda|
           |\varphi_b^2-\varphi_b+\lambda|}\,d\varphi_b.
\label{eq:l}
\end{eqnarray}

The coefficients in the continuum quantum superposition I shall take to be the real positive square roots of this measure.  I should like to emphasize that, like all other proposals for the quantum state of the universe, this is just a proposal and is not derived from previously accepted principles.

The symmetric-bounce proposal specifies the form of the quantum state at the bounce, but, unlike some other proposals such as the symmetric initial condition \cite{CZ}, it does not impose any requirement that the wavefunction be normalizable over the entire superspace.  Indeed, even for the minisuperspace of the homogeneous isotropic modes of the scale factor variable $\alpha$ and the inflaton field variable $\varphi$, the symmetric-bounce wavefunction propagates unabated to arbitrarily large $\alpha$ and so is not normalizable, that is, it is not square-integrable over the $(\alpha,\varphi)$ space with the area element induced from the DeWitt metric \cite{DeW}.

Because the symmetric-bounce hypersurface
$(\alpha_{\mathrm{bounce}}(\varphi_b),\varphi_b)$ becomes asymptotically
null sufficiently rapidly with $|\varphi_b|$ for large $|\varphi_b|$, so
that $\mu(\varphi_b) \sim |\varphi_b|^{-3/2}$ for large $|\varphi_b|$,
the total measure $\mu(\varphi_b)d\varphi_b$ integrated over all
$\varphi_b$ from minus infinity to plus infinity is finite.  It is
dominated by the regions where $\varphi_b^2 \sim \lambda$, giving
$\int_{-\infty}^{\infty}\mu(\varphi_b)d\varphi_b \approx
(4/3)\lambda^{-3/4} \approx 7\times 10^{82}$ for $\lambda \approx
5.0\times 10^{-111}$ as estimated above.  Here I shall ignore one-loop
quantum corrections \cite{BK90,BKK92,BK94}, partly because of the fact
that if they are important, unknown higher-loop effects are likely also
to be important.  Such quantum corrections should be unimportant when the energy density is much less than the Planck density, e.g., for $\varphi^2 \ll G^{-1}m^{-2} \sim 10^{12}$.  The energy density at the bounce is less than the Planck value for over 99.9\% of the measure of the symmetric bounce trajectories with $\varphi_b^2 > 1$.

The symmetric-bounce homogeneous spacetimes, labeled by the value of
$\varphi_b$ where each of them has its symmetric bounce on the
symmetric-bounce hypersurface $(\alpha_{\mathrm{bounce}}(\varphi_b),
\varphi_b)$, may be divided into five classes depending on which
spacelike or timelike segment of the symmetric-bounce hypersurface at
which each of them has its symmetric bounce.  These segments are
divided by the points at which the symmetric-bounce hypersurface
becomes null in the DeWitt metric of Eq.\ (\ref{eq:c}) and crosses
from being spacelike to timelike or from timelike to spacelike.  These
points are where $1-[d\alpha_{\mathrm{bounce}}(\varphi_b)/d\varphi_b]^2
= 0$ or $(\varphi_b^2+\varphi_b+\lambda)
(\varphi_b^2-\varphi_b+\lambda) \equiv (\varphi_b+\varphi_2)
(\varphi_b+\varphi_1) (\varphi_b-\varphi_1) (\varphi_b-\varphi_2)=0$,
or at $\varphi_b = -\varphi_2$, $\varphi_b = -\varphi_1$, $\varphi_b =
+\varphi_1$, and $\varphi_b = +\varphi_2$, where $\varphi_1 =
(1/2)(1-\sqrt{1-4\lambda}) \approx \lambda$ and $\varphi_2 =
(1/2)(1+\sqrt{1-4\lambda}) \approx 1$.  Then one may define Segment 1
to be the spacelike part of the symmetric-bounce hypersurface with
$\varphi_b < -\varphi_2$, Segment 2 to be the timelike part with
$-\varphi_2 < \varphi_b < -\varphi_1$, Segment 3 to be the spacelike
part with $-\varphi_1 < \varphi_b < \varphi_1$, Segment 4 to be the
timelike segment with $\varphi_1 < \varphi_b < \varphi_2$, and Segment
5 to be the spacelike segment with $\varphi_2 < \varphi_b$.  Under the
symmetry $\varphi \rightarrow -\varphi$, Segments 1 and 5 are
interchanged, Segments 2 and 4 are interchanged, and Segment 3 is
interchanged with itself.  Therefore, without loss of generality, one
may take $\varphi_b \geq 0$ and consider only Segments 3, 4, and 5.  One may estimate that for $\lambda \approx 5.0\times 10^{-111}$, Segments 1 and 5 each have measure $\approx (1/2)B(1/4,3/2) \approx 1.748$, Segments 2 and 4 each have measure $\approx (2/3)\lambda^{-3/4} \approx 3.5\times 10^{82}$, and Segment 3 has measure $\approx (\pi/2)\lambda^{1/4} \approx 4.2\times 10^{-28}$.

At a symmetric bounce, using the gauge $n=1$, one has $\dot{\alpha} =
\dot{\varphi} = 0$, but $\ddot{\alpha} =
e^{-2\alpha_{\mathrm{bounce}}(\varphi_b)} = \varphi_b^2 + \lambda$ and
$\ddot{\varphi} = -\varphi_b$, so the trajectory starts with the slope
$d\alpha/d\varphi = \ddot{\alpha}/\ddot{\varphi} = -(\varphi_b^2 +
\lambda)/\varphi_b = d\varphi_b/d\alpha_{\mathrm{bounce}}(\varphi_b)$,
orthogonal to the symmetric-bounce hypersurface in the DeWitt metric of
Eq.\ (\ref{eq:c}).  As one moves slightly away from the symmetric
bounce, $\varphi$ always starts evolving toward zero, and $\alpha$
always starts evolving toward larger values.  For a symmetric bounce in
Segments 1, 3, and 5, the trajectory initially moves into the
minisuperspace region above the symmetric-bounce hypersurface, $\alpha >
\alpha_{\mathrm{bounce}}(\varphi)$, and is there timelike in the DeWitt
metric ($d\alpha^2 > d\varphi^2$); for a symmetric bounce in Segments 2
and 4, the trajectory initially moves into the minisuperspace region
below the symmetric-bounce hypersurface, $\alpha <
\alpha_{\mathrm{bounce}}(\varphi)$, and is there spacelike in the DeWitt
metric ($d\alpha^2 < d\varphi^2$).

Symmetric-bounce homogeneous spacetimes that bounce on Segment 3
thereafter move along timelike trajectories ever upward in the
$(\alpha,\varphi)$ minisuperspace and hence expand forever.  Their
dynamics are always dominated by the positive cosmological
constant and behave very nearly like empty de Sitter universes.  In my proposed measure, their measure is only $\sim 10^{-28}$ that of Segments 1 and 5 and only $\sim 10^{-110}$ that of Segments 2 and 4, so these nearly empty spacetimes do not seem to contribute much to the measure for observations, unlike their contribution to the Hartle-Hawking no-boundary quantum state \cite{Susspriv,DKS,GKS,Sus03,DP2006}.

Symmetric-bounce spacetimes that bounce on Segment 2 or 4, with
$\lambda^2 \approx \varphi_1^2 < \varphi_b^2 < \varphi_2^2 \approx 1$,
except for $\varphi_b^2$ sufficiently close to 1, generally have a
period of expansion during which the scalar field oscillates rapidly
relative to the expansion.  When averaged over each oscillation, the
mean value of $\dot{\varphi}^2$ is nearly the same as that of
$\varphi^2$ (in a gauge with $n=1$, which I shall assume unless stated
otherwise), which is equivalent to saying that the pressure exerted by
the scalar inflaton averages to near zero over each oscillation.  Then
the scalar field acts essentially like pressureless dust, with a total
rationalized dimensionless `mass' that is nearly constant:
\begin{equation}
M \equiv (\varphi^2 + \dot{\varphi}^2)r^3
 = (\varphi^2 + \dot{\varphi}^2)e^{3\alpha} = \frac{8\pi G}{3}m\rho a^3,
\label{eq:m}
\end{equation}
where $a=r/m$ is the physical scale factor and
\begin{equation}
\rho = 
\frac{1}{2}\left[m^2\phi^2
+\left(\frac{1}{N}\frac{d\phi}{dt}\right)^2\right]
=\frac{3m^2}{8\pi G}(\varphi^2 + \dot{\varphi}^2)
\label{eq:m2}
\end{equation}
is the energy density of the scalar field with our choice of $n=mN=1$ to
make our time coordinate $t$ dimensionless (and with $d/dt$ being
denoted by an overdot).  Thus the dimensionless $M$ is $4Gm/(3\pi)$
times the integral of the energy density $\rho$ over the volume $2\pi^2
a^3$ of the 3-sphere of physical scale factor $a$ and of dimensionless
scale factor $r \equiv e^{\alpha} \equiv ma$.  The approximate constancy
of $M$ during the `dust' regime results from the fact that the integral
of
\begin{equation}
\frac{dM}{d\alpha} = 3(\varphi^2 - \dot{\varphi}^2)e^{3\alpha}
\label{eq:n}
\end{equation}
is approximately zero over each oscillation of the scalar field.

Then during such a `dust' phase, the dimensionless scale factor $r=ma$
evolves according to
\begin{equation}
\dot{r}^2 = \lambda r^2 + \frac{M}{r} - 1
\label{eq:o}
\end{equation}
with $M$ very nearly constant.  As a function of the dimensionless scale
factor $r \equiv ma$ at fixed $M$, the right hand side has a minimum at
$r = [M/(2\lambda)]^{1/3}$ that is positive if $27\lambda M^2 > 4$, so
when this condition holds, the universe will expand forever from any
initial $r$ if $M$ stays constant.  However, this sufficient (but not
necessary) condition for expansion forever does not hold for any
$\varphi_b^2 \ll 1$ for which $M$ stays nearly constant after the
bounce, at which one has

\begin{equation}
r_b = \frac{1}{\sqrt{\varphi_b^2+\lambda}},\ 
M_b = \frac{\varphi_b^2}{(\varphi_b^2+\lambda)^{3/2}},
\label{eq:p}
\end{equation}
since obviously the right hand side of Eq.\ (\ref{eq:o}) is zero at
the bounce.

That is, although $27\lambda M^2 > 4$ with constant $M$ is sufficient
for the universe to expand forever in our simple $k=+1$ FRW model with a
cosmological constant and a massive scalar field that acts like dust, it
is not necessary.  Conversely, $27\lambda M^2 < 4$ is necessary but not
sufficient for recollapse.  If $27\lambda M^2 < 4$ does hold, one also
needs that $r$ be at an allowed value (one giving $\dot{r}^2 \geq 0$)
less than the minimum of the right hand side of Eq.\ (\ref{eq:o}), which
is equivalent to $2\lambda r^3 < M$.  Thus this model $k=+1$ FRW
$\Lambda$-dust model will recollapse (assuming $M$ stays constant) if and only if
\begin{eqnarray}
&&2\lambda r^3 < M < \frac{2}{\sqrt{27\lambda}} \Leftrightarrow
   \nonumber \\
&&27\lambda^3 r^6 < 6.75\lambda M^2 < 1.
\label{eq:q}
\end{eqnarray}

Using Eq.\ (\ref{eq:p}), which leads to a nearly constant $M \approx
M_b$ when $\varphi_b^2 \ll 1$, we see that our $k=+1$ FRW
$\Lambda$-scalar model with the symmetric-bounce initial condition
will recollapse if and only if
\begin{equation}
10^{-110} \approx 2\lambda \stackrel{<}{\sim} \varphi_b^2
\stackrel{<}{\sim} O(1).
\label{eq:r}
\end{equation}
This is the part of Segments 2 and 4 with larger values of
$\varphi_b^2$, plus a bit into Segments 1 and 5. For $\lambda \ll
\varphi_b^2$, well into the interior of this open set of values of $\varphi_b$, the evolution will have $\lambda r^2 \ll M/r$ during the evolution, so the dimensionless collapse time $\Delta t$ with $n=1$ will be approximately $(\pi/2)r_b \approx \pi/(2\varphi_b)$.  For $\varphi_b^2$ large enough to give a density sufficient for nucleosynthesis (e.g., at the density our universe had at an age of a few minutes), the lifetime in proper time would be of the order of minutes, far too short for the evolution of stars and observers that depend upon stars.  Although Segments 2 and 4 dominate the measure given by Eq.\ (\ref{eq:l}) by factors of the order of $10^{82}$, they do not do so by factors anywhere near the inverses (say $\sim e^{10^{42}}$) of the exponentially tiny relative probabilities of forming Boltzmann brains, so the resulting symmetric-bounce universes will presumably have extremely tiny probabilities for observers and should contribute negligibly to observational probabilities.  (This is unlike the case of the Hartle-Hawking no-boundary proposal, where factors from the negative Euclidean action, say $\sim e^{10^{122}}$, {\it can} be much greater than the inverses of the relative probabilities to form Boltzmann brains or even Boltzmann solar systems.)

The symmetric-bounce initial conditions that lead to recollapse
actually extend past $\varphi_b^2 = \varphi_2^2 \approx 1$ into
Segments 1 and 5, but there the dust approximation that $M$ is nearly
constant breaks down.  It is difficult to give a good approximate
closed-form treatment for $\varphi_b^2 \sim 1$, but for $\varphi_b^2$
a few times unity, one enters the slow-roll inflationary regime where
$M$ grows greatly during a period of inflation that can be estimated
fairly accurately under the approximation that $\varphi_b^2 \gg 1$. 

\section{Approximate solutions for the inflationary regime}

Let us now focus on the regime in which the initial (at the bounce)
value of $\varphi^2 \equiv 4\pi G \phi^2/3$, that is $\varphi_b^2$, is
at least somewhat large compared to unity, so that the evolution away
from the symmetric bounce starts with a period of slow-roll inflation
that includes at least several e-folds of expansion.  In this Section,
we want to set up some theoretical analysis before turning in the next
Section to a numerical calculation of how many e-folds of inflation
occur, as a function of $\varphi_b$, and also of the
$\varphi_b$-dependence of the asymptotic value, in the `dust' regime
that follows the inflationary regime, of the total rationalized
dimensionless `mass' $M$ given by Eq.\ (\ref{eq:m}).

Without loss of generality, assume that the value of $\varphi$ at the
bounce, $\varphi_b$, is positive, so when it is greater than $\varphi_2
\approx 1$, the FRW spacetime starts on Segment 5 with $r_b \approx
1/\varphi_b$.  During the slow-roll inflationary regime with $\varphi
\gg 1$, we have $\varphi^2 \gg \dot{\varphi}^2$.  Since during inflation
we have $\dot{\varphi}^2 + \varphi^2 \stackrel{>}{\sim} 1 \gg \lambda$,
we can neglect the cosmological constant term $\lambda$ during inflation
and take the inflationary equations to be Eqs.\ (\ref{eq:g}) or
(\ref{eq:h}) with $\lambda$ dropped:
\begin{eqnarray}
&&\dot{r}^2 = r^2(\varphi^2 + \dot{\varphi}^2) - 1,
   \nonumber \\
&&\ddot{\varphi} + 3\frac{\dot{r}}{r}\dot{\varphi} + \varphi = 0,
\label{eq:s}
\end{eqnarray}
in terms of the dimensionless scale factor $r \equiv e^\alpha \equiv
ma$, or
\begin{eqnarray}
&&\dot{\alpha}^2 = \varphi^2 + \dot{\varphi}^2 - e^{-2\alpha},
   \nonumber \\
&&\ddot{\varphi} + 3\dot{\alpha}\dot{\varphi} + \varphi = 0,
\label{eq:t}
\end{eqnarray}
in term of the logarithm $\alpha = \ln(ma)$ of the scale factor, and in terms of the dimensionless form $\varphi \equiv \sqrt{4\pi G/3}\phi$ of the inflaton scalar field $\phi$.

From Eqs.\ (\ref{eq:s}), one can readily derive, as an alternative to
the redundant Eq.\ (\ref{eq:h2}) when $\lambda$ is neglected, that
\begin{equation}
\ddot{r} = r(\varphi^2-2\dot{\varphi}^2). 
\label{eq:u}
\end{equation}
We shall define the inflationary period as that first period immediately
after the symmetric bounce when $\lambda$ is negligible (so as not to
consider inflation by the cosmological constant) and when $\ddot{r} >
0$, so that the scale factor of the universe is accelerating with
respect to cosmic proper time.  This is equivalent, with $\lambda$
negligible, to the first period during which $2\dot{\varphi}^2 <
\varphi^2$.  Let us define $N$ (or $N(\varphi_b)$, since it depends on
the initial value $\varphi_b$ at the bounce) to be the number of
e-folds of the inflationary period, the change in the logarithm $\alpha$
of the scale factor during the inflationary period that starts with
$\varphi = \varphi_b > 0$ and $\dot{\varphi}=0$ at $\alpha =
\alpha_b(\varphi_b) \equiv \alpha_{\mathrm{bounce}}(\varphi_b) =
-\ln{\varphi_b}$ by Eq.\ (\ref{eq:k}) with $\lambda$ neglected and that
ends at $\alpha = \alpha_e(\varphi_b)$ where $\varphi$ has first dropped
to the then-positive value of $-\sqrt{2}\dot{\varphi}$:
\begin{equation}
N(\varphi_b) \equiv \alpha_e(\varphi_b) - \alpha_b(\varphi_b). 
\label{eq:v}
\end{equation}

It also is convenient to define a shifted scale-factor logarithm
\begin{equation}
\beta \equiv \alpha - \alpha_b \equiv \alpha + \ln{\varphi_b}, 
\label{eq:w}
\end{equation}
which increases monotonically from $\beta_b = 0$ at the bounce to
$\beta_e = N$ at the end of the inflationary period.  Then the
$\varphi_b$-dependent number of e-folds of inflation may be defined to
be $N(\varphi_b) = \beta_e(\varphi_b)$.  $N(\varphi_b)$ will be large if
$\varphi_b \gg 1$, which is what we shall assume, though many of the
results below turn out to be quite accurate even if $\varphi_b$ is as
small as 3.

Now I shall give a sequence of increasingly better approximations for
the early phase of inflation, followed by numerical calculations of
$N(\varphi_b)$ and of the aftermath of inflation, such as the asymptotic
value of the total rationalized dimensionless `mass' $M$ given by Eq.\
(\ref{eq:m}).

The simplest approximation is for the period when $\varphi$ remains very
nearly the same as its initial value $\varphi_b$ and when
$\dot{\varphi}$ is negligible in comparison.  Then the first of Eqs.\
(\ref{eq:s}) becomes $\dot{r}^2 \approx r^2\varphi_b^2 - 1$, with the
solution
\begin{equation}
r \approx \varphi_b^{-1}\cosh{\varphi_b t}, 
\label{eq:x}
\end{equation}
which gives de Sitter spacetime at this level of approximation. 
However, this level of approximation does not remain good indefinitely,
since the second of Eqs.\ (\ref{eq:s}) implies that $\varphi$ gradually
decreases.

For $\varphi_b t \gg 1$ but still $\varphi^2 \gg \dot{\varphi}^2$ (so
that several e-folds of inflation have occurred but one is not yet near
the end of inflation), one is in the flat ($e^{-2\alpha} \ll \varphi^2 +
\dot{\varphi}^2$) slow-roll ($\dot{\varphi}^2 \ll \varphi^2$) regime
where the first of Eqs.\ (\ref{eq:s}) or (\ref{eq:t}) now becomes
$\dot{r} \approx r\varphi$ or $\dot{\alpha} \approx \varphi$, so that
the second of Eqs.\ (\ref{eq:s}) or (\ref{eq:t}) becomes $\ddot{\varphi}
+ 3\varphi\dot{\varphi} + \varphi \approx 0$, which has the attractor
solution \cite{Star78}
\begin{equation}
\varphi = {\mathrm{const.}} - t/3 \sim \varphi_b -t/3. 
\label{eq:x2}
\end{equation}
Then one gets 
\begin{equation}
\alpha \approx {\mathrm{const.'}} + ({\mathrm{const.}})t - t^2/6
\sim \alpha_b + \varphi_b t -t^2/6 \sim \alpha_b +
1.5(\varphi_b^2-\varphi^2). 
\label{eq:x3}
\end{equation}
Since inflation ends when $\varphi$ drops down to
$-\sqrt{2}\dot{\varphi}$, which by the slow-roll approximation (no
longer valid near the end of inflation but giving the right order of
magnitude) is $\sqrt{2}/3$, which is much less than $\varphi_b$ that we
are assuming is much larger than unity, we get as the leading
approximation for the number of e-foldings of inflation that
$N(\varphi_b) \sim 1.5 \varphi_b^2$.  However, we shall find below that
there is also a term logarithmic in $\varphi_b$, as well as terms that
are inverse powers of $\varphi_b^2$, plus a constant term that may be
evaluated numerically.

If one looks at just the flat regime where $r^2(\varphi^2 +
\dot{\varphi}^2) \gg 1$ but does not impose the slow-roll condition
$\dot{\varphi}^2 \ll \varphi^2$, one can see that Eq.\ (\ref{eq:i}) with
$U \equiv -\alpha' \equiv -d\alpha/d\varphi$ becomes the autonomous
first-order differential equation
\begin{equation}
\frac{dU}{d\varphi} = (U^2 - 1)(\frac{U}{\varphi}-3). 
\label{eq:y}
\end{equation}
During slow-roll inflation with $\varphi \gg 1$, the solution will
exponentially rapidly approach the attractor solution
\begin{equation}
U = 3\varphi + \frac{1}{3\varphi} - \frac{2}{27\varphi^3}
   +\frac{11}{243\varphi^5} - \frac{10}{243\varphi^7} + O(\varphi^{-9}).
\label{eq:z}
\end{equation}
This then gives
\begin{equation}
\alpha \approx {\mathrm{const.}} - \frac{3}{2}\varphi^2 -
   \frac{1}{3}\ln{\varphi} - \frac{1}{27\varphi^2}
   +\frac{11}{972\varphi^4} - \frac{5}{729\varphi^6} + O(\varphi^{-8}), 
\label{eq:aa}
\end{equation}
where the const.\ term depends upon $\varphi_b$. One can see that this
formula leads to a $(1/3)\ln{\varphi_b}$ term in $N(\varphi_b)$, but
the value of the constant term in $N(\varphi_b)$ and of the terms that
go as inverse powers of $\varphi_b^2$ require the behavior both before
the entry into the flat regime and after the exit from the slow-roll
regime.

Next, let us go to a better approximation during the first stages of
inflation, not assuming one has entered the flat regime where the
spatial curvature term $e^{-2\alpha}$ may be neglected.  If one inserts
the approximate solution for $r(t)$ from Eq.\ (\ref{eq:x}) into the
second one of Eqs.\ (\ref{eq:s}) and solves it to the leading nontrivial order in $1/\varphi_b$, one gets the better approximation for the scalar field that is (cf.\ \cite{Star96})
\begin{equation}
\varphi \approx \varphi_b 
- \frac{1}{3\varphi_b}[\ln{\cosh{(\varphi_b t)}}
+\tanh^2{(\varphi_b t)}]. 
\label{eq:bb}
\end{equation}
Analogously, if one inserts the approximate solution for $\varphi(t)$
from Eq.\ (\ref{eq:x2}) into the first one of Eqs.\ (\ref{eq:s}) and
solves it under the slow-roll approximation, one gets the better
approximation for the dimensionless scale factor $r=ma$ that is
\begin{equation}
r \equiv e^\alpha \approx \varphi_b^{-1}\cosh{(\varphi_b t-t^2/6)}. 
\label{eq:cc}
\end{equation}

Both of these approximations are valid for all $t \ll \varphi_b$, both
the regime in which the spatial curvature is not negligible and the
early stages of the slow-roll regime in which $\varphi$ has not rolled
down very close to the bottom.  One might have thought it would be yet a
better improvement to take the argument of the hyperbolic functions in
the expression for $\varphi$ to be the same as they are given in the
hyperbolic functions in the expression for $r$, namely $\varphi_b
t-t^2/6$, but this would invalidate the fact that during the entire
flat slow-roll regime, $\dot{\varphi}$ stays very close to $-1/3$.
For $1 \ll \varphi_b t \ll \varphi_b^2$, so that one is in the early
part of the flat slow-roll regime, one has
\begin{equation}
\varphi \approx \varphi_b - \frac{1}{3} t - \frac{1-\ln{2}}{3\varphi_b} 
\label{eq:dd}
\end{equation}
and
\begin{equation}
\alpha \equiv \ln{r} \approx \varphi_b t
 - \frac{1}{6}t^2 - \ln{(2\varphi_b)}
\approx \frac{3}{2}\varphi_b^2 -\ln{\varphi_b}
-\frac{3}{2}\varphi^2-(1-\ln{2})\frac{\varphi}{\varphi_b}-\ln{2}. 
\label{eq:ee}
\end{equation}

For an even better approximation during the early stages of the
slow-roll regime, one can use Eq.\ (\ref{eq:j}) and the definition $R
\equiv e^\beta = r/r_b = \varphi_b r = \varphi_b m a$ to get
\begin{eqnarray}
\varphi \approx && 
\varphi_b-\frac{1}{3\varphi_b}\left[\ln{R}+1-\frac{1}{R^2}\right]
\nonumber \\
&-& \frac{1}{162\varphi_b^3}[9\ln^2{R}+12\ln{R}
      -\frac{54}{R^2}\ln{R}
\nonumber \\
&+&18\left(\arccos{\frac{1}{R}}\right)^2
+36\left(\arccos{\frac{1}{R}}\right)\frac{\sqrt{R^2-1}}{R^2}
\nonumber \\
&-&16+\frac{3}{R^2}+\frac{9}{R^4}+\frac{4}{R^6}]
+O(\varphi_b^{-5}).
\label{eq:ff}
\end{eqnarray}
Taking this expression into the flat regime for which $\beta \equiv
\ln{R} \gg 1$ gives
\begin{equation}
\varphi \approx \varphi_b - \frac{\beta+1}{3\varphi_b}
-\frac{9\beta^2+12\beta+4.5\pi^2-16}{162\varphi_b^3}+O(\varphi_b^{-5}). 
\label{eq:gg}
\end{equation}

When this approximation for $\varphi(\beta)$ in the flat slow-roll
regime is inverted and matched to Eq.\ (\ref{eq:aa}), one gets
\begin{equation}
\alpha \approx \frac{3}{2}\varphi_b^2 - \frac{2}{3}\ln{\varphi_b}
-\frac{3\pi^2-14}{36\varphi_b^2}
-\frac{3}{2}\varphi^2 - \frac{1}{3}\ln{\varphi} -\frac{1}{27\varphi^2},
\label{eq:hh}
\end{equation}
neglecting uncalculated terms going as higher inverse powers of
$\varphi_b^2$ and both calculated and uncalculated terms going as higher
inverse powers of $\varphi^2$.  From this expression, one can see that
at the end of inflation,
\begin{equation}
\alpha_e \approx \frac{3}{2}\varphi_b^2 - \frac{2}{3}\ln{\varphi_b}
-\frac{3\pi^2-14}{36\varphi_b^2} + {\mathrm{const.}}
\label{eq:ii}
\end{equation}
and the number of e-folds of inflation is
\begin{equation}
N(\varphi_b) = \beta_e = \alpha_e -\alpha_b \approx
\frac{3}{2}\varphi_b^2 + \frac{1}{3}\ln{\varphi_b}
-\frac{3\pi^2-14}{36\varphi_b^2} + {\mathrm{const.}},
\label{eq:ii2}
\end{equation}
but, so far as I can see, the numerical constant in this expression
cannot be determined by a closed-form expression but requires numerical
integration to the end of inflation at $\varphi =
-\sqrt{2}\dot{\varphi}$, which is beyond the validity of the slow-roll
approximation used above that applies for $\varphi \gg
-\sqrt{2}\dot{\varphi}$.

\section{Numerical results for the inflationary regime}

Since the closed-form approximate expressions derived above do not apply
near the end of the inflationary regime, I used Maple to get fairly
precise numerical expressions of how many e-folds $N(\varphi_b)$ of
inflation occur (the increase in the logarithmic scale factor $\alpha =
\ln{(ma)}$ during the inflationary period that is defined as the initial
period during which the second time derivative of the scale factor,
$\ddot{a}$, is positive), and of what the asymptotic value
$M_\infty(\varphi_b)$ of the dimensionless `mass' $M$ is, as functions
of the initial value $\varphi_b$ of the dimensionless inflaton scalar
field $\varphi \equiv \sqrt{4\pi G/3}\phi$ here written in terms of the
physical inflaton scalar field $\phi$.

I integrated the equations of evolution from the bounce to the end of
inflation for several values of $\varphi_b$  and found that for $\varphi_b \stackrel{>}{\sim} 10$,
\begin{equation}
N(\varphi_b) \approx \frac{3}{2}\varphi_b^2
+ \frac{1}{3}\ln{\varphi_b} -1.0653 
-\frac{3\pi^2-14}{36\varphi_b^2} -\frac{0.4}{\varphi_b^4}.
\label{eq:mm}
\end{equation}

I also found that at the end of inflation for $\varphi_b \stackrel{>}{\sim} 3$, $\varphi \approx 0.4121$ and $\dot{\varphi} \approx -0.2914$, about one-eighth of the way from its slow-roll value of $-1/3$ to zero.  From this one can also deduce that at the end of inflation, $M = M_e \approx 0.2547\,\varphi_b^{-3} e^{3N(\varphi_b)}$.

The next question is the $\varphi_b$-dependent value of
$M_\infty(\varphi_b)$, the asymptotic value of the total rationalized
dimensionless `mass' $M = (\varphi^2 + \dot{\varphi}^2)r^3 = (8\pi
G/3)m\rho a^3$, where $\rho$ is the scalar field energy density.  To
make the definition precise, one could take $M_\infty$ to be the value
of $M$ at infinite time if the cosmological constant is positive and if
the solution expands forever, and to be the value of the dimensionless
scale factor $r=ma$ (or, more precisely, of $r(1-\lambda r^2)$ if
$\lambda$ were not negligible as it is in practice) at the first maximum
of $r$ if the universe does not expand forever (which will necessarily
be the case if the cosmological constant is not positive).  However, in
practice, the dimensionless cosmological constant $\lambda \equiv
\Lambda/(3m^2) \approx 5.0\times 10^{-111}$ is so tiny that it is
insignificant during the numerical integrations of the inflationary
regime, and for large $\varphi_b$ the maximum $r \sim \exp{(4.5
\varphi_b^2)}$ before the universe would recollapse in the absence of a
positive cosmological constant is so huge that one cannot take the
numerical integrations that far.  Therefore, I shall approximate
$M_\infty(\varphi_b)$ by the value $M$ settles down toward in the `dust'
regime after the end of inflation but long before one needs to consider
the effects of either $\lambda$ or the spatial curvature $e^{-2\alpha}$.

Numerically, it is still a bit tricky to get precise values for
$M_\infty(\varphi_b)$, because $M(t)$ oscillates along with $\varphi$
(at twice the frequency and at the harmonics of that frequency, since
$M(t)$ depends only on $\varphi^2(t)$ and $\dot{\varphi}^2(t)$), with
oscillation magnitudes of the basic frequency and its harmonics that
decay only as inverse powers of the scale factor.  However, one can
derive that the following function eliminates the first several
harmonics and after the end of inflation rapidly settles down very near
its asymptotic value $M_\infty(\varphi_b)$:
\begin{eqnarray}
M_{\mathrm{asym}}(t) = e^{3\alpha}\{(\varphi^2+\dot{\varphi}^2)
+3\dot{\alpha}\varphi\dot{\varphi}
+\frac{9}{32}\left[9(\varphi^2+\dot{\varphi}^2)^2-8\dot{\varphi}^4
+\dot{\alpha}\varphi\dot{\varphi}(3\varphi^2+\dot{\varphi}^2)\right]
   \nonumber \\
+\frac{81}{128}\left[10(\varphi^2+\dot{\varphi})^3
-15\varphi^2\dot{\varphi}^4-11\dot{\varphi}^6
+5\dot{\alpha}\varphi\dot{\varphi}
(6\varphi^4+4\varphi^2\dot{\varphi}^2-3\dot{\varphi}^4)\right]\}.
\label{eq:kk}
\end{eqnarray}

My numerical results gave
\begin{equation}
M_\infty(\varphi_b) \approx 0.1815\varphi_b^{-3}e^{3N(\varphi_b)}
\approx \frac{0.08914 e^{4.5\varphi_b^2}}
{12\varphi_b^2+3\pi^2-14+24/\varphi_b^2}.
\label{eq:nn}
\end{equation}
One can see that $M_\infty(\varphi_b) \approx 0.7125 M_e$, 71\% of the
value of $M$ at the end of inflation, because of the decaying
oscillations of $M(t)$ after the end of inflation.

One can now use this formula along with the criterion of the rightmost inequality of Eq.\ (\ref{eq:q}) to deduce that for inflationary solutions starting on Segment 5 with $\lambda = 5\times 10^{110}$, one needs $\varphi_b \stackrel{>}{\sim} 5.4646$ or $\phi_b \stackrel{>}{\sim} 2.6700\,G^{-1/2}$ or $N(\varphi_b) \stackrel{>}{\sim} 44.28$ e-folds of inflation to avoid eventual recollapse and instead have expansion forever in an asymptotic de Sitter regime.  This is assuming that the simple inflaton-$\Lambda$ model applied for all time.  In a more realistic model in which the energy of the inflaton field converted to radiation shortly after the end of inflation, one would need a larger initial inflaton field value $\varphi_b$ and more e-folds of inflation to avoid eventual collapse.  For example, if one had all the energy of the inflaton field convert to radiation right at the end of inflation and the universe evolve thereafter as a radiation-$\Lambda$ model thereafter, one would need $16\lambda M_e r_e > 1$, which by using the formula above for $M_\infty(\varphi_b)$ and the relation $M_\infty(\varphi_b) \approx 0.7125 M_e$ gives $\varphi_b \stackrel{>}{\sim} 6.6069$ or $\phi_b \stackrel{>}{\sim} 3.2282\,G^{-1/2}$ or $N(\varphi_b) \stackrel{>}{\sim} 65.03$ to avoid eventual recollapse.  It is rather remarkable that despite the extremely tiny value of $\lambda$, the critical initial values of the inflaton field $\phi_b$ are within one-half an order of magnitude of being unity, essentially because of the very rapid growth of $M_\infty(\varphi_b)$ with $\varphi_b$.

Another asymptotic constant late in the `dust' regime (but before either
the cosmological constant term $\lambda$ or the spatial curvature term
$e^{-2\alpha}$ becomes important) is the asymptotic value of a certain
phase $\theta$.  At late times in the `dust' regime, ignoring $\lambda$
and $e^{-2\alpha}$, one can write $\varphi=\dot{\alpha}\cos{\psi}$ and
$\dot{\varphi}=-\dot{\alpha}\sin{\psi}$ to define an evolving phase
angle $\psi$, and then the asymptotically constant phase is
\begin{equation}
\theta = \frac{2}{3\dot{\alpha}}-\psi+\sin{\psi}\cos{\psi}.
\label{eq:oo}
\end{equation}
(There are more complicated formulas that I have derived for the
asymptotically constant phase in the de Sitter phase and/or when the
spatial curvature is not negligible, but I shall leave them for a later
paper.)  Preliminary numerical calculations suggest that the asymptotic
value of $\theta$, say $\theta_\infty(\varphi_b)$, is roughly 1.978 for
large $\varphi_b$, but I have not had time to confirm this and to
investigate the dependence on $\varphi_b$.

For solutions of our system of a $k=+1$ Friedmann-Robertson-Walker universe with a minimally coupled massive scalar field and a positive cosmological constant that have a bounce at a minimal value of the scale factor and then expand forever in an asymptotically de Sitter phase, there will be an analytic map (not known explicitly, of course) from the initial values at the bounce of $\varphi$ and $\dot{\varphi}$, say $\varphi_b$ and $\dot{\varphi_b}$, to the asymptotic values $M_\infty(\varphi_b,\dot{\varphi_b})$ and $\theta_\infty(\varphi_b,\dot{\varphi_b})$ (or more precisely, to $M_\infty(\varphi_b,\dot{\varphi_b})$ and the complex constant $C_\infty(\varphi_b,\dot{\varphi_b}) = e^{i\theta_\infty(\varphi_b,\dot{\varphi_b})}$, since $\theta_\infty(\varphi_b,\dot{\varphi_b})$ is actually only defined modulo $2\pi$, but for simplicity I shall continue to refer to $\theta_\infty(\varphi_b,\dot{\varphi_b})$).  For the symmetric-bounce solutions, the solution-space is just one-dimensional (governed by the one parameter $\varphi_b$) rather than two-dimensional, with the restriction $\dot{\varphi_b}=0$, so both $M_\infty$ and $\theta_\infty$ are then functions just of $\varphi_b$.  Hence for these symmetric-bounce
solutions, in principle one gets a particular analytic relation
$\theta_\infty = \theta_{\infty,{\mathrm{sb}}}(M_\infty)$.

For the complex solutions of the same minisuperspace system
corresponding to the no-boundary proposal \cite{HHH}, one should get a
slightly different analytic relation $\theta_\infty =
\theta_{\infty,{\mathrm{nb}}}(M_\infty)$, though one would expect these
two functions to approach the same values for very large $M_\infty$.  In
the no-boundary case, in which the one free parameter is the complex
initial value of $\varphi$, say $\varphi(0)$, both
$M_\infty(\varphi(0))$ and $\theta_\infty(\varphi(0))$ would be complex
for generic complex $\varphi(0)$, but one could choose a
one-real-parameter contour in the complex-$\varphi(0)$ plane that would
make $M_\infty(\varphi(0))$, say, real.  But it would still be the case
that even for real $M_\infty(\varphi(0))$, the corresponding
$\theta_\infty(\varphi(0))$ would not be quite real, so
$\theta_{\infty,{\mathrm{nb}}}(M_\infty)$ would not be precisely real
for real $M_\infty$ as $\theta_{\infty,{\mathrm{sb}}}(M_\infty)$ is for
the symmetric-bounce solutions, as it always is for a real one-parameter
set of Lorentzian spacetimes of the FRW form being assumed here. 
Therefore, it is a bit ambiguous what real Lorentzian solutions
correspond to the no-boundary proposal, even asymptotically, since for
the complex extrema obeying the no-boundary conditions, one cannot have
the two asymptotic constants $M_\infty$ and $\theta_\infty$ both real. 
One can of course make {\it ad hoc} choices, such as taking the real
Lorentzian solutions that corresponds to real values of $M_\infty$ and
then to the real values $\theta_\infty =
\mathrm{Re}(\theta_{\infty,{\mathrm{sb}}}(M_\infty))$ that are the real
parts of the complex values $\theta_{\infty,{\mathrm{sb}}}$ given by the
no-boundary proposal for the real values of $M_\infty$. However, one
does need to make some such {\it ad hoc} choice before getting precisely
real Lorentzian solutions from the no-boundary proposal.

\section{Inhomogeneous and/or anisotropic perturbations}

The symmetric-bounce proposal for the quantum state of the universe is
that the universe has inhomogeneous and anisotropic quantum
perturbations about the set of classical inflationary solutions
described above that are in their ground state at the symmetric bounce
hypersurface.  In particular, the quantum state of the perturbations on
that hypersurface is proposed to be the same as that of the de
Sitter-invariant Bunch-Davies vacuum \cite{Bunch-Davies} on a de Sitter
spacetime with the same radius of the throat as that of the classical
background symmetric-bounce inflationary solution at its throat.

Of course, once the massive scalar inflaton field starts to roll down
its quadratic potential, the background spacetime will deviate from de
Sitter spacetime, so that the quantum perturbations will no longer
remain in a de Sitter-invariant state.  One would expect the usual
inflationary picture of parametric amplification that would result in
each inhomogeneous mode leaving its initial vacuum state and becoming
excited as the wavelength of that mode is inflated past the Hubble scale
given by the expansion rate.  In this way one would get the usual
inflationary production of density perturbations arising from the
initial vacuum fluctuations.

This part of the story is similar to the Hartle-Hawking no-boundary
proposal \cite{HVat,HH,H,HalHaw,P,Hal,HLL,HH1,Page-Hawking-BD,HH2,HHH},
which also predicts that the inhomogeneous and anisotropic quantum
perturbations start off in the de Sitter-invariant Bunch-Davies vacuum
(and admittedly predicts this in a slightly less {\it ad hoc} way than
it is proposed in my symmetric-bounce proposal).  However, the main
difference is that the symmetric-bounce proposal has the more uniform
weighting given by Eqs.\ (\ref{eq:l}) for the different values of
$\varphi_b$ and hence of the dimensionless bounce radius $r_b =
1/\varphi_b$, rather than being weighted by the exponential of twice the
negative action of the Euclidean hemisphere as in the no-boundary
proposal.  It is this exponential weighting of the no-boundary proposal
that apparently leads to the probabilities being enormously dominated by
the largest Euclidean hemispheres, those of empty de Sitter spacetime,
and hence for observational probabilities dominated by early-time
Boltzmann brains (or Boltzmann solar systems, if one excludes the
possibility of observers existing without an entire solar system)
\cite{Susspriv,DKS,GKS,Sus03,DP2006}.  By not having these Euclidean
hemispheres and their enormously negative Euclidean actions, the
slightly more {\it ad hoc} symmetric-bounce proposal can avoid the huge
domination by empty or nearly-empty de Sitter spacetimes that seems very
strongly at odds with our observations of significant structure far
beyond ourselves, such as stars.

\section{Conclusions}

The symmetric-bounce proposal is that the quantum state of the universe
is a pure state that consists of a uniform distribution (in a metric induced from the DeWitt metric on the superspace) of components (of different bounce sizes) that each have the quantum fluctuations initially (at the bounce) in their ground state at a moment of time symmetry for a bounce of minimal three-volume.  The background spacetimes of this proposal (ignoring the quantum fluctuations) consist of a one-parameter family (at least for one inflaton field; if there are more, there would be as many parameters as bounce values of all the inflaton fields) of time-symmetric inflationary Friedmann-Robertson-Walker universes.  For each member of this family, the quantum state of the inhomogeneous and/or anisotropic fluctuations are, at the bounce, the same as the de Sitter-invariant Bunch-Davies vacuum for a de Sitter spacetime with the same curvature as the background FRW universe at its bounce.  The entire quantum state is a coherent superposition of all these FRW spacetimes with their quantum fluctuations, with weights given by the DeWitt metric for the bounce configurations.

This symmetric-bounce quantum state reproduces all the usual predictions
of inflation but avoids the huge negative Euclidean actions of the
Hartle-Hawking no-boundary proposal that seems to make the probabilities
dominated by nearly-empty de Sitter spacetime and make our observations
of distant structures (e.g., stars) extremely improbable.

It is interesting that since the background inflationary FRW cosmologies
for each macroscopic component of the symmetric-bounce quantum state are
time symmetric about a bounce, there is actually no big bang or other
initial singularity in this model.  The classical background universes
contract down to the bounce without becoming singular, and then they
re-expand in a time-symmetric way.  However, because the quantum
fluctuations are in their ground state at the bounce, that is the moment
of minimal coarse-grained entropy, so entropy grows away from the bounce
in both directions of time.  Any thermodynamic observer would sense that
the arrow of time (given by the observer's memories and observations of
the increase of entropy) is increasing away from the bounce, so it would
regard the bounce as in its past.  Thus one would get the observed time
asymmetry of the universe without any of the background classical
components having this asymmetry in a global sense.  In Wheeleresque
terms, the universe would have time-asymmetry without time-asymmetry.

\section*{Acknowledgments}

I appreciated the hospitality of the Mitchell family and of the George P.\ and Cynthia W.\ Mitchell Institute for Fundamental Physics and Astronomy of Texas A\&M University at a workshop at Cook's Branch Conservancy, where the basic idea for this paper arose while I was kayaking around Firemeadow Lake, and where I had many useful discussions on related issues with the workshop participants, especially on this particular issue with Jim Hartle and Thomas Hertog.  I am thankful to an anonymous referee for suggesting many improvements and added references, and to Bill Unruh and the University of British Columbia for hospitality while these corrections were made.  This research was supported in part by the Natural Sciences and Engineering Research Council of Canada.

\end{document}